\documentclass[aps,prb,superscriptaddress,amsmath,amssymb,preprintnumbers,twocolumn,floatfix,showpacs,showkeys,10pt]{revtex4}
\usepackage{amssymb}
\usepackage{epsfig}

\begin{document}
\title{Quantum teleportation  of electrons in quantum wires 
with surface acoustic waves}
\author{Fabrizio~Buscemi }
\email{fabrizio.buscemi@unimore.it}
\affiliation{ARCES, Alma Mater Studiorum, University of Bologna, Via Toffano 2/2, 40125 Bologna, Italy}
\affiliation{S3 Research Center, CNR-INFM, Via Campi 213/A, I-Modena 41125, Italy}
\author{Paolo~Bordone}
\affiliation{S3 Research Center, CNR-INFM, Via Campi 213/A, I-Modena 41125, Italy}
 \affiliation{ Dipartimento di Fisica, Universit\`{a} di Modena e
 Reggio Emilia, 41125, Modena, Italy}
\author{Andrea~Bertoni }
\affiliation{S3 Research Center, CNR-INFM, Via Campi 213/A, I-Modena 41125, Italy}

\begin{abstract}
We propose and numerically simulate a semiconductor device based
on coupled quantum wires, suitable for  deterministic quantum
teleportation of electrons trapped in the minima of surface acoustic
waves. We exploit a network of interacting semiconductor quantum wires able to
provide the universal set of gates for quantum information processing,
with the qubit defined by the localization of a single electron in one
of two coupled channels.
The numerical approach is based on a time-dependent solution of the
three-particle Schr\"odinger equation.  First, a maximally entangled pair
of electrons is obtained via Coulomb interaction between
carriers in different channels.  Then, a complete Bell-state
measurement involving one electron from this pair and a third electron
is performed.  Finally, the teleported state is reconstructed by means
of local one-qubit operations. The large estimated fidelity explicitely 
suggests that an efficient
teleportation process could be reached in an experimental setup.

\end{abstract}

\maketitle

\section{Introduction} 
Quantum teleportation is the process where a quantum state is
transferred from one system to another one at a different location.
It relies on quantum entanglement, the most peculiar trait of quantum
mechanics.  In the protocol described by Bennett \emph{et
al.}\cite{PhysRevLett.70.1895} the sender Alice and the receiver Bob
share an entangled pair of particles.  Alice entangles her particle
with a third one, namely the one whose state is to be teleported.
Then she performs a destructive joint-measurement on the two-particle
system on her side.  Next, she communicates through a classic channel
the outcome of the measure to Bob.  Through this information he can
reconstruct the original quantum state by simply applying local
operations on his particle.
Clearly the scheme of teleportation relies on the prior establishment
of quantum entanglement between the two parties.  However, only
classical communication is used after the particle to be teleported
comes into play at Alice side.

While experimental realization of quantum teleportation 
protocols has
been performed in NMR~\cite{Nielsen}, optical~\cite{Zhao}, and
atomic systems~\cite{S.Olmschenk01232009}, no evidence of
teleportation in semiconductor systems has been achieved so far.
Indeed, semiconductor technology represents a viable approach for 
the realization of quantum computing devices, and quantum teleportation
would be a crucial validation of its potentiality. Theoretical proposals 
for electron teleportation in solid-state
systems are based on single~\cite{Wang,Sauret} 
and double~\cite{deVisser,dePasquale,Reina} quantum dots. 
Teleportation protocols using edge channels in the
quantum Hall effect have  also been  advanced~\cite{Beenakker}. 
However, quantum-wire systems have the advantage of intrinsically
providing the qubit transmission between specified  locations, as
required by the DiVincenzo criteria~\cite{DiVincenzo}.  
Furthermore, they could
be directly integrated in traditional electronic circuitry and allow
in principle the implementation of a large number of quantum hardware
units thus overcoming the scalability problem.
In this frame, coherent electron transport in systems of couples of
semiconductor quantum wires has been used to design qubits and to
propose fundamental one- and two-qubit quantum gates~\cite{Bertoni,Ion}.
Furthermore, the use of surface acoustic waves (SAWs) as a mean to
inject and drive carriers along the wires presents some advantages
with respect to the free propagation along quasi-1D channels since it
prevents the spreading of the electron wavefunction, it reduces
undesired reflection effects, and it makes the electron more immune to
the decohering effects of the
phonons~\cite{Buscemi,Rodriquez,Barnes1}.

In this work we propose and simulate numerically a scheme to perform
quantum deterministic teleportation of electrons in a device 
consisting of three
couples of semiconductor quantum wires.  The carriers are
embedded in the minima of SAWs, propagating in the wires direction.
The qubit-state is encoded through the localization of a single
electron in one of two parallel quantum wires.
In our scheme, the Coulomb interaction between carriers is used first
for the production of an Einstein-Podolsky-Rosen (EPR) pair of 
electrons, and then for the
rotation of the Bell states needed to perform a Bell
measurement\cite{PhysRevLett.70.1895,Benenti}.
We note that the teleportation model described in the following could
also be applied, without qualitative modifications, to edge channels
in the quantum Hall regime.   In fact, the latter system has already
successfully exposed the two-particle quantum interference via an
electronic version of the Hanbury Brown Twiss
setup~\cite{Samuelsson,Neder}.

\section{The physical system} \label{Phy}
The physical system used to 
implement our quantum teleportation scheme
consists of three electrons injected by SAW along three couples of
GaAs quantum wires.  We assume that the device operates at low
temperatures (simulations are performed at zero temperature) in order
to have a negligible number of electrons in the conduction band and to
minimize  decoherence effects due to the interactions of electron with
lattice vibrations.
The use of SAW for the injection and transport of electrons in the
quantum wires~\cite{Barnes1} allows a single carrier to be captured
into a minimum of the sinusoidal piezoelectric wave propagating along
the device and inhibits the natural spatial spreading of the wavepacket.
In this way the particle is confined in a moving quantum dot~\cite{Shilton} and a
so-called flying qubit is realized~\cite{Rodriquez}.

In order to implement the quantum operations for 
the teleportation scheme we employ three elements: an electronic beam
splitter $R_x(\theta)$, a phase shifter $R_{0(1)}(\phi)$ and a Coulomb coupler
$T(\gamma)$~\cite{Barenco}.

The former is realized through a coupling window between
the two wires of a qubit, able to split an incoming wavefunction into
two parts~\cite{bird}. In terms of qubit transformations it corresponds to
$R_x (\theta)|0\rangle = \cos{(\theta/2)}|0\rangle+ i\sin{(\theta/2)}|1 \rangle$ and
$R_x (\theta)|1\rangle = i\sin{(\theta/2)}|0\rangle+ \cos{(\theta/2)}|1 \rangle$.
The electronic phase shifter $R_{0(1)}(\phi)$ is realized by introducing a
suitable potential barrier in the wire $0$($1$).
It acts only on a single qubit state by adding a phase factor, namely
$R_{0} (\phi)|0\rangle = e^{i\phi}|0\rangle$ and $R_{0}
(\phi)|1\rangle = |1\rangle$. Similarly, $R_{1} (\phi)|0\rangle =|0\rangle$ and $R_{1} (\phi)|1\rangle =  e^{i\phi}|1\rangle$.
The $T(\gamma)$ Coulomb coupler is the only two-qubit gate.  It consists of a
region in which two electrons propagating along two different wires
get close enough to each other to give rise to an effective
interaction.  A phase $\gamma$ is added if and only if the two-qubit
systems is in $| 0 \rangle |1\rangle $ and the $T(\gamma)$ gate acts
as follows: $T(\gamma) | 0 \rangle |1\rangle =
e^{i\gamma} |0 \rangle |1 \rangle $, leaving the other three two-qubit
states unchanged.  The above quantum gates have been numerically
validated elsewhere~\cite{Bertoni2}.
In fact, the phases $\theta$, $\phi$ and $\gamma$ depend upon the 
physical and geometrical parameters of the system, e.g. the window
length, the barrier height/length, the electron-electron coupling
strength, the SAW velocity.  The proposed teleportation scheme
(with the exception of  qubit 3 preparation)   employs only quantum
gates with $\theta=\pi/2$, $\phi=\pi$,  $\gamma=\pi$ and we tuned the
device parameters accordingly.  For brevity, in the following we will
omit the indication of the above phases.

\begin{figure}[h]
  \begin{center} 
  \includegraphics*[width=\linewidth]{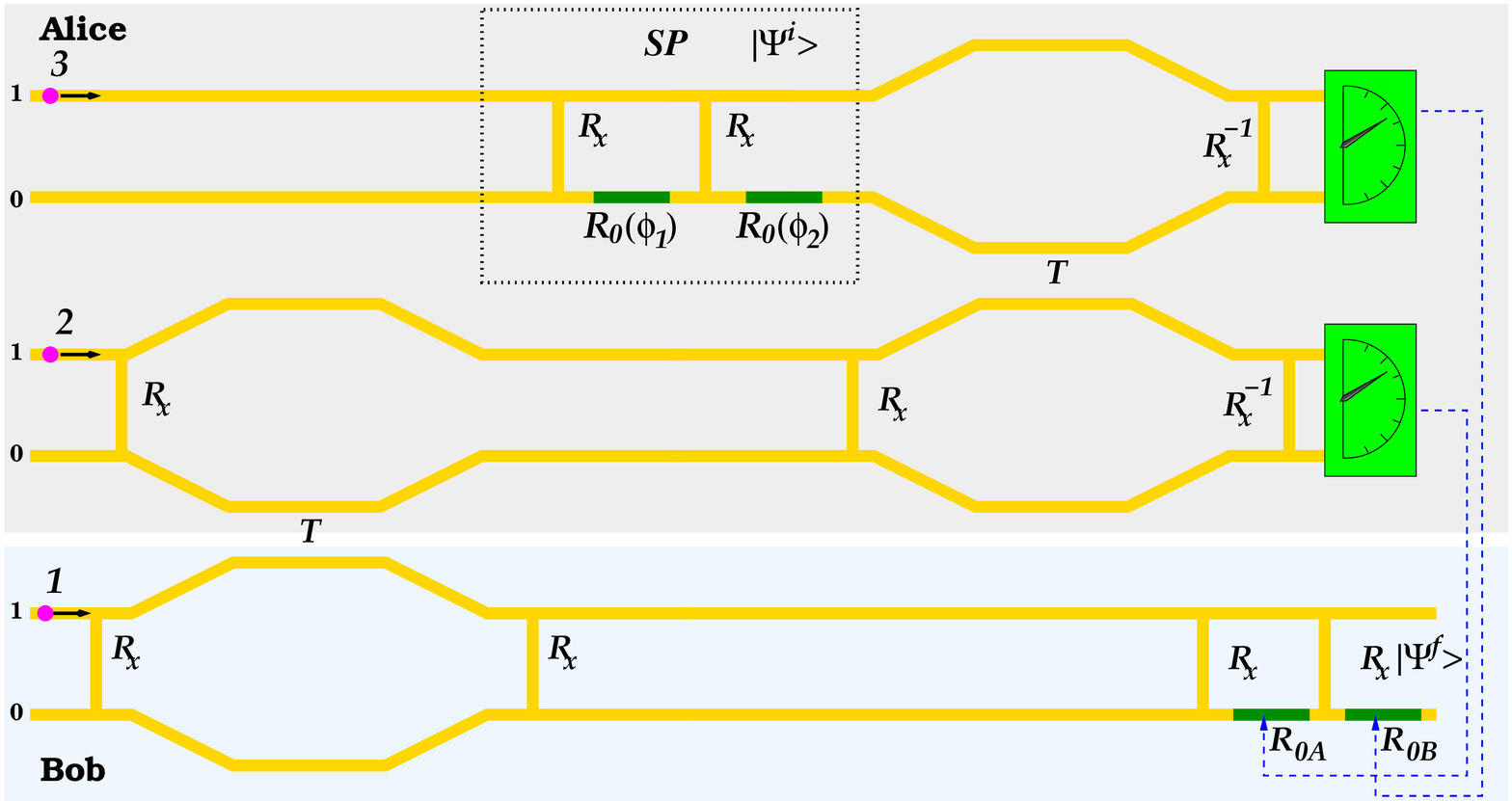}
  \caption{\label{fig1} (Color online) Sketch of the physical system used for the
  deterministic teleportation of electron 3 in electron 1. The first 
two couples of quantum wires from the top represent the Alice's system, 
while the bottom couple is Bob's system.  The Bell-state preparation of qubits
 1 and   2 consists of two beam splitters $R_x^{(1)}$ and $R_x^{(2)}$  
followed by a Coulomb coupler $T^{(12)}$ and a further
  splitting $R_x^{(1)}$ of qubit 1. The gates in the dotted box labelled SP 
is needed to prepare the input state $|\Psi_3^{i} \rangle$.  This block applied
  to $|1_3\rangle$ produces the general  one qubit-state
  $|\Psi_3^{i} \rangle=e^{i\phi_2} \cos{(\phi_1/2)} |0\rangle -
  \sin{(\phi_1/2)} |1\rangle $.  The Bell-measurement process in
  Alice's system is realized in two sequential steps.  First a
  complete rotation from Bell states to separable states is performed
  by means of the beam splitter  $R_x^{(2)}$ followed by a Coulomb coupler
  $T^{(23)}$ and further inverse rotations of $\pi/2$
  ${R_x^{-1}}^{(2)}$ and ${R_x^{-1}}^{(3)}$. Then the single-qubit states
  are measured by means of charge detectors (green boxes). The outcome is 
communicated through a   classical channel (dashed lines) to Bob, 
which reconstructs in  $|\Psi_1^{f} \rangle$ the
  original quantum state of qubit 3  by means of a network of one-qubit
  gates (two beam splitters $R_x^{(1)}$ and 
  two potential barriers). Specifically, the potential barrier $R_{0A}$
($R_{0B}$) is switched on if and only if the outcome of qubit 2(3) 
measurement is 0 (see Table~\ref{tab1}).}
  \end{center}
\end{figure}

The quantum-wire network for our teleportation protocol is shown in
Figure~\ref{fig1}.  As a first step Alice and Bob must entangle 
qubits 2 and 1.  In particular, they start with an initial
state $|1_2 1_1\rangle$~\cite{Note1},corresponding to one
electron entering the upper wire of Bob qubit 1 and another electron
entering the upper  wire of Alice qubit 2.  The first block of quantum
gates, namely two coupling windows $R_x$ (acting on qubits 1 and 2,
respectively) and a Coulomb coupler $T$ (acting on qubits 1 and 2,
together) produces a maximally entangled state $1/\sqrt{2}\left(|0_2
1_1\rangle-|1_2 0_1\rangle\right)$.  Now Alice, 
wants to teleport,   the quantum state of electron 3 $|\Psi_3^i\rangle=
s^i |0_3\rangle +t^i|1_3\rangle$, obtained from $|1_3\rangle$ by means  
of the  network of one-qubit gates reported in  the State Preparation (SP) 
box, into electron 1 at  Bob side. She performs a so-called Bell
measurement on qubits 3 and 2.  In fact, as suggested by Brassard
\emph{et al.}~\cite{Brassard}, such a measurement can be
realized in two steps: first the Bell states of qubits 2 and 3 are
rotated in the basis $(|0_3 0_2\rangle, |0_3 1_2\rangle, |1_3
0_2\rangle, |1_3 1_2\rangle)$, then a projective measurement is
performed in this latter basis.  This approach permits to achieve
deterministic quantum teleportation, since it makes possible a full
Bell measurement on Alice's particles.  In our scheme, the first step
of the Brassard approach is performed by means of the second block of
quantum gates involving a Coulomb coupler $T$ now acting on the qubits 2
and 3 and three $R_x$ gates (see Fig.~\ref{fig1}).    The
 three-qubit state obtained after this block takes the form:

 \begin{eqnarray} \label{outgoing}
& &\!\!\!\!\!\!\!\! \!\! |\Phi_{3,2,1}\rangle_{OUT}= {}  \nonumber \\
& & \!\!\!\!\!\!\!\! \!\!-\frac{1}{2}|0_30_2 \rangle \big(s^f|0_1\rangle\! +\!
 t^f|1_1\rangle \big)
\! +\!\frac{1}{2}|0_31_2 \rangle \big(t^f|0_1\rangle\! -\! s^f|1_1\rangle \big)  \nonumber  \\
& &\!\!\!\!\!\!\!\! \!\!-\frac{i}{2}|1_30_2 \rangle\big(\!\!-\!\!s^f|0_1\rangle\!+\!t^f|1_1\rangle \big)\!+\!\frac{i}{2}|1_31_2 \rangle \big(t^f|0_1\rangle\!+\!s^f|1_1\rangle \big) 
 \end{eqnarray}
After this rotation Alice can perform two single-particle measurements
on qubits 2 and 3.  Specifically, such measurements can be realized by
means of single-electron transistors, acting as sensitive charge
detectors.  Once the result is known, it can be transmitted as two
classical bits of information to Bob, that can choose the setup of
proper unitary operations on his qubit 1 in order to completely
recover the initial state $|\Psi^i _3\rangle$. In fact, depending
upon the outcome of the measurement on qubits 2 and 3, suitable
potential barriers acting as phase shifters $R_0$ are eventually introduced
between the two coupling windows, according Table~\ref{tab1}. 

\begin{table}
\begin{tabular}{|c|c|c|c|}
  \hline
  qubit 3 & qubit 2 & ${R_0}_B$ & ${R_0}_A$ \\  \hline  \hline 
  0 & 0 & YES  & YES \\  \hline
  0 & 1 & YES     & NO \\  \hline
  1 & 0 & NO  & YES\\  \hline
  1 & 1 & NO &  NO  \\  \hline
\end{tabular}
\caption{ The phase shifters (potential barriers) in the final stage of the Bob qubit are applied according to the outcome of Alice measurements in order to reconstruct the original state $|\Psi^i _3\rangle$. Note that the  phase shifter ${R_0}_B({R_0}_A)$ is controlled only by the qubit 3 (2), i.e.  the  potential barrier  is introduced only when the electron is found in the lower wire of the corresponding qubit at Alice side.  
\label{tab1} }
\end{table}


\section{The numerical approach}\label{numapproach}
The network of gates described in Sec.~\ref{Phy} has been simulated by solving
numerically the time-dependent Schr\"odinger equation for the three
electrons injected in the device.  GaAs material parameters have been
used.  Since a direct solution of the 3D-Schr\"odinger equation for
the whole three-particle wavefunction results too demanding in terms
of computational resources, a semi-1D model has been adopted, as
described in the following by referring to Fig.~\ref{fig1}.  The
network is defined in the $xy$-plane.  In the $z$ direction the
electrons are always supposed to be in the ground state of the quantum
well defining the plane of the wires.  The $y$ direction is
explicitely included in the simulations through the $y_1,y_2,y_3$
variables, defining the position of the three carriers along the
wires. Specifically, in the computational approach adopted
$y$ is discretized with a point-grid of resolution $\Delta y =1$nm.  For the $x$ direction, the $x_1,x_2,x_3$ variables can assume
only the values 0 or 1, identifying one of the two possible wires of a
qubit, that is the qubit state. 

This allows us to move from a time dependent Schr\"odinger equation
for a seven-variable wavefunction $\Phi(x_1,x_2,x_3, y_1,y_2,y_3, t) $
to eight coupled Schr\"odinger equations of the form
\begin{eqnarray} \label{Sch}
&&i\hbar \frac{\partial}{\partial t}\Phi_{x_1,x_2,x_3}(y_1,y_2,y_3,t)= \nonumber \\
{} & &-\frac{\hbar^2}{2m}\left(\frac{\partial^2}{\partial y_1^2}
+ \frac{\partial^2}{\partial y_2^2} + \frac{\partial^2}{\partial y_3^2}\right)
\Phi_{x_1,x_2,x_3}(y_1,y_2,y_3,t) \nonumber \\ 
& &+ V_{x_1,x_2,x_3}(y_1,y_2,y_3,t) \Phi_{x_1,x_2,x_3}(y_1,y_2,y_3,t) . 
\end{eqnarray}
The potential term appearing in the above equation
is given by the sum of two contributions.  The first 
stems from the SAW time-dependent potential and reads
\begin{eqnarray}
 \sum^3_{i=1} A \sin{\left[\frac{2\pi}{\lambda}(y_i- v_s t)\right]}\qquad \quad \qquad \forall \{x_1,x_2,x_3\},
\end{eqnarray}
where $A$ indicates the  amplitude of the potential, $\lambda$  its  wavelength,
and  $v_s$ the sound velocity. Specifically, in the numerical investigations performed, $A$= 20 meV, $\lambda$= 200 nm, and  $v_s= 3.3 \times 10^5$ cm s$^{-1}.$  The second term  represents  the screened Coulomb interaction between carriers, computed from the geometry of the system and can be written  in the form 
\begin{eqnarray}
  \sum^3_{i=1} \sum^{i-1}_{j=1} \frac{e^2}{4\pi \epsilon_0 r_{ij}} \exp{ \left(-\frac{r_{ij}}{r_0}\right)},
\end{eqnarray}
where $r_{ij}$=$\sqrt{(y_i-y_j)^2+d_{x_i,x_j}^2(y_i,y_j)}$ with $d_{x_i,x_j}(y_i,y_j)$ indicating
the distance between the  wires $x_i$ and $x_j$ at the positions  $y_i$ and
$y_j$, respectively.  This is  a Coulomb potential multiplied by an exponential damping term, corresponding to Debye  wave vector $1/r_0$. In particular, in our calculations the latter has been  taken equal to 0.2 nm$^{-1}$: a value of the same of order of the ones given in the literature~\cite{Ferry}. Due to the  geometry of the system and to screening effects, the  Coulomb interaction  can be considered  negligible everywhere for not adjacent qubits, that is  qubit 1 and 3, and in the regions not involving the Coulomb coupler  for nearby qubits.

Each of  the eight coupled Schr\"odinger equations of  Eq.~(\ref{Sch}) has  been solved by means of simple-finite-difference relaxation method applied at each time step of the time evolution performed in a Crank-Nicholson scheme with $\Delta t$=0.01 fs~\cite{NumRec,Note2}.

It is  worth noting that  in the simulation of the three-particle wave function dynamics, the 
only gate that presents a computational challenge due to its spatial
extension and the two-particle potential involved, is the Coulomb
coupler $T$.  However, it is not always necessary to compute its
effect on the whole wave function.  In fact, when the first $T$ gate 
comes into action by entangling electrons 1 and 2, electron 3 wave
function remains factorizable. As a consequence, before
the second $T$ gate, a two-particle simulation is sufficient.

The numerical estimation of the second $T$ gate, between qubit 2 and
3, requires more care since at this stage qubit 1 is already entangled
with qubit 2. This entanglement is not only related to the 
localization of electrons in one of their wires, but also to their positions $y_1$
and $y_2$.  As a consequence, the effect of the second Coulomb coupler
must be computed on the whole three-particle wave function.  Thanks to
the superposition principle and to the fact that the interaction is
just among electrons 2 and 3, we performed different two-particle
simulations for different values of $y_1$ and then computed the final
state as the combination of the different evolutions.  We found that,
due to the sharp localization of the spatial wavepackets and to the
small $y$-direction entanglement, the solution turns out to be practically
independent from the choice of $y_1$, as it will be shown by the numerical results reported in the next section.

A number of numerical simulations have  been performed  in order to obtain the optimal geometry for the Coulomb coupler $T$. This  is reached when the  delay phase $\gamma$ attains  $\pi$. As shown in other works~\cite{Bertoni3}, the latter mainly depends upon two geometrical  parameters: the length of the coupling region and the distance between the coupled wires. From the optimization procedure,  we find that the Coulomb coupler is  150 nm long while the coupled wires   are  5 nm distant  from each other. This geometry  allows   a value of  0.88$\pi$ for the delay-phase $\gamma$, which is  good enough for realizing  both the initial  maximally entangled state and the final rotation of the  Bell states at Alice side. The  experimental realization of  the Coulomb coupler $T$ is the most challenging part.  Specifically,  the angle formed by a wire  where it bends towards the other qubit  must be small enough  in order to make  reflection phenomena negligible. In addition,  no tunnelling  between the two wires must be present to let the two wavefunctions only interact through Coulomb coupling.

Obviously, within our semi 1D-model it is not possible to simulate
directly the dynamics of the wave function splitting by a coupling
window leading to the one-qubit transformation $R_x$.  In fact, we
exploited the results of 2D simulations to validate the qubit
transformations~\cite{Bertoni} and include the beam splitters through
their corresponding transformation matrix.

\section{Results and discussion}
  We have performed our numerical simulations to
teleport many  test states prepared by tuning the phase $\phi_1$ and
setting $\phi_2$ to $\pi/2$ for the phase shifters in the SP box of
Fig.~\ref{fig1}.

\begin{figure}[h]
  \begin{center} \includegraphics*[width=\linewidth]{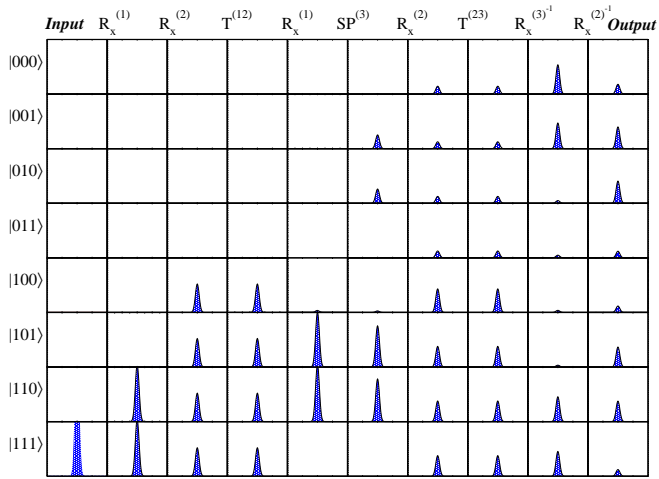}
  \caption{\label{fig2} (Color online) The effect of the quantum gates of the network
  of Fig.~\ref{fig1} on a three-qubit state at different stages of the
  time evolution.  The square modulus  of the eight components of three-carrier wavefunction   $\Phi(\bar{y},\bar{y},\bar{y})$ are reported as a function of $\bar{y}$ at different time   steps. Here the teleported
  state is  $|\Psi^i_3\rangle= (1/2)|0_3\rangle +(i\sqrt{3}/2)|1_3\rangle$. Thus the
output state is $|\Phi_{3,2,1}\rangle_{OUT}$ of  Eq.~(\ref{outgoing})
  with   $s^f=(1/2)$ and $t^f=(i\sqrt{3}/2)$. Note that, to optimize the
  graphical represention, the curves are normalized to the ones
  corresponding to the states $|1_3 1_2 0_1\rangle$ and $|1_3 1_2
  1_1\rangle$ in the second column.}  \end{center}
\end{figure}

Figure~\ref{fig2} summarizes the three-qubit dynamics estimated
numerically for the case of
$|\Psi^i_3\rangle= (1/2)|0_3\rangle +(i\sqrt{3}/2)|1_3\rangle$ 
(corresponding to $\phi_1= (2\pi/3)$), starting from the
carrier-injection instant up to the single-particle measurements on
Alice's qubits.  The square modulus of the
eight components of the three-particle wavefunction
$\Phi_{x_1,x_2,x_3}(\bar{y},\bar{y},\bar{y},t)$ is reported
being the latter evaluated for the three
electrons  in the same $\bar{y}$ position.
We initialized the system in  $|1_3 1_2 1_1\rangle$ (electrons injected
in the upper wire of each couple).  The first column of
Fig.~\ref{fig2} shows the only non vanishing component of the
wavefunction.
The two $R_x$ gates located in the left part of the device split 
in the same way the electrons of qubits 1 and 2
(third column of Fig.~\ref{fig2}).
Then, the Coulomb coupler acting between
qubits 1 and 2 induces a phase of 0.88$\pi$ in  $|1_3 0_2 1_1\rangle$ 
with respect to the other
components. 
When the injected carriers reach the second coupling
window between the wires of qubit 1, the new rotation leads with good
approximation to the three-qubit state $(1/\sqrt{2})|1_3 \rangle
(|0_2 \rangle|1_1 \rangle-|1_2 \rangle| 0_1
\rangle)$, as shown in the fifth column.
This corresponds to an EPR pair of  electrons 1 and 2.  However, we
find that the component   $|1_3 0_2
0_1\rangle$ is small but not zero.  This can be ascribed to the fact
that the rotation performed by the simulated $T$ gate is not exactly
$\pi$. Then the single-qubit gates of the SP block operate onto $|1_3 \rangle$
and  the state $(1/\sqrt{2})|\Psi_3^i \rangle (|0_2 \rangle|1_1 \rangle-|1_2 \rangle| 0_1 \rangle)$ is produced.

 From this stage, single and two-qubit operations act only onto
electrons 2 and 3, in order to perform the complete rotation of the Bell state
of qubits 2 and 3 into separable states.  
The components of the three-particle wavefunction displayed in the
last column of Fig.~\ref{fig2} show the state
$|\Phi_{3,2,1}\rangle_{OUT}$ on which the destructive measurements
will be performed by Alice.  The state components depend on the
coefficients of the spectral decomposition of the teleported state
$|\Psi_1^f\rangle$ in terms of the single qubit-states $|0_1\rangle$
and $|1_1\rangle$.  In agreement with the theoretical prediction, we
find that, for $|\Psi_3^i\rangle= (1/2)|0_3\rangle
+(i\sqrt{3}/2)|1_3\rangle$, the square modulus of the components $|0_3
0_2 1_1\rangle$, $|0_3 1_2 0_1\rangle$, $|1_3 0_2 1_1\rangle$, and
$|1_3 1_2 0_1\rangle$ has the same form and value, which approximately
is the triple of the one found for $|0_3 0_2 0_1\rangle$,$|0_3 1_2
1_1\rangle$, $|1_3 0_2 0_1\rangle$, and $|1_3 1_2 1_1\rangle$,
respectively.  Such a result shows the good efficiency reachable in
the proposed teleportation scheme.

\begin{figure}[h]
  \begin{center}
    \includegraphics*[width=\linewidth]{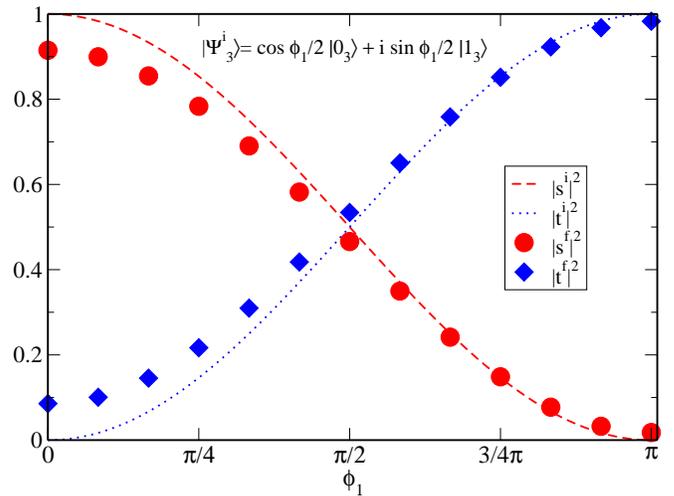}
   \caption{\label{fig3} (Color online) Square modulus of the coefficients of  the initial state $|\Psi^i_3\rangle=s^i |0_3\rangle +t^i |1_3\rangle$ and the final state $|\Psi^f_1\rangle=s^f |0_1\rangle +t^f |1_1\rangle$  as a function of $\phi_1$.
This graph illustrates the case with  $s^i=\cos{(\phi_1/2)}$ and  $t^i=i\sin{(\phi_1/2)}$. Note that the phase $\phi_1$ is tuned by the barrier height in the $R_1$  gate of the SP box of Fig.~\ref{fig1}.}
 \end{center}
\end{figure}

To better quantify the reliability of the teleportation, we also
compare the square modulus of the coefficients $s^i$ and $t^i$ of
 the initial state $|\Psi^i_3\rangle$ with the coefficients $s^f$ and $t^f$ of the
final  state $|\Psi^f_1\rangle$ obtained by Bob after the
teleportation (see Fig.~\ref{fig3}). The fidelity $F$ of the
teleportation process, given by $\left|\langle
  \Psi^{i}_3|{\Psi}_{1}^f\rangle\right|^2$, is strictly related to the
ratio $|{s^i}|^2/|{s^f}|^2$: the closer to 1 the latter is, the
larger values the fidelity attains. For the set of teleported states,
obtained by varying the phase $\phi_1$ from 0 to $\pi$, the above
ratio ranges from $0.91$ $(F=0.91)$ to 1.02 $(F=0.98)$ and, for
$\phi_1=(3\pi/4)$, it is almost equal to 1 $(F=1)$.  This implies that
the state $\cos{\left[\left( 3\pi/8\right)\right]} |0\rangle +
i\sin{\left[\left( 3\pi/8\right)\right]}$ is teleported with the
maximum efficiency.  This proves an important point: the fidelity of
the proposed teleportation scheme remains very high also for  non
ideal entangling gates and this can certainly 
be considered a plus in view of an experimental
implementation.

\begin{figure}[h]
  \begin{center}
    \includegraphics*[width=\linewidth]{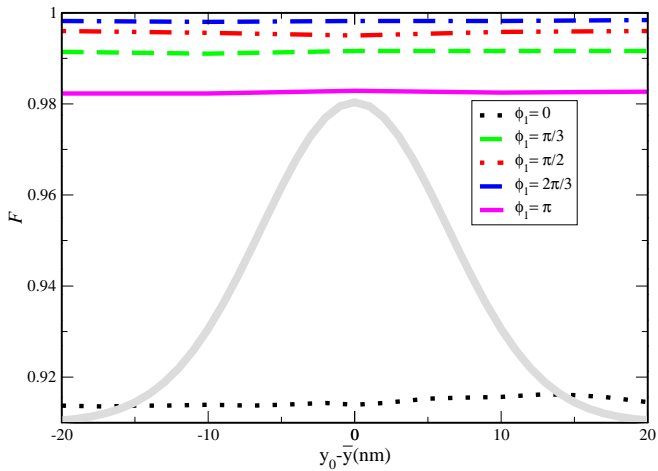}
      \caption{\label{fig4} (Color online) Fidelity of the teleportation process as a function of the difference between the wavepacket center $y_0$ and the point  $\bar{y}$, for five teleported  states corresponding to different values of the phase $\phi_1$ with  $\phi_2=\pi/2$: $\phi_1=0$ (dotted line), $\phi_1=\pi/3$ (dashed line), $\phi_1=\pi/2$ (dot-dot-dashed line), $\phi_1=2\pi/3$ (dash-dash-dotted  line), and $\phi_1=\pi$ (solid line).   The abscissa scale, ranging   from -20 nm to 20 nm,  covers the space region where the integral of the single-electron probability density (shown for reference and represented by the Gaussian-like  thick solid line) is equal to 0.92. Note that the estimated  fidelities  take values  very close to  1 and  that,for any teleported state, do not  significantly depend upon position in the examined space interval.}
 \end{center}
\end{figure}

In order to test the validity  of the approach used 
in the numerical solution of  the 3D-Schr\"odinger equation for three-particle 
wavefunction,  we have reported in  Fig.~\ref{fig4}, the fidelity $F$ as a function of  $\bar{y}$ (that is the point along the wires of qubit 3 where the carrier is assumed to be found in a measurement process). This is repeated for five teleported states
corresponding to different values of the phase  $\phi_1$ with  $\phi_2$ set to 
$\pi/2$. We find that $F$  is essentially constant.  This implies that the efficiency of our quantum teleportation scheme does not  basically depend  upon  the position variable along the wire direction of  the three particles.  This behavior,  due to the confinement of  the carriers in the SAW minima leading to a sharp localization of the corresponding  wavepackets,  proves  the validity of the two-particle approach adopted.

Finally, the effect  of the temperature deserves a few comments. As stated in Sec.~\ref{Phy},  numerical simulations of  charge transport through quantum wires has been performed at zero temperature, that is fully coherent propagation of electrons has been assumed. Such an approximation allows one to neglect the  effects due to the piezoelectric coupling between charge carriers and acoustic phonon modes. In fact, the latter represents  the main decoherence source of our physical system.   Experimental investigations about both SAW assisted charge transport~\cite{Naber} and low-dimensional devices suitable for quantum computing~\cite{Yang}  are usually carried out at temperatures in the range of mK. A  realization of our device would require  such low temperatures at which  the stimulated absorption and emission processes of acoustic phonons  are weak enough to be neglected. In fact,   the mean occupation number of acoustic phonons with  momentum energy $E_k\approx$ 5 meV (of the order of the energy
difference between the ground and the first excited bound state of the electrons) can be evaluated from Bose-Einstein statistics and results to be practically zero  at a temperature of 100 mK. Thus it seems  reasonable to take into account only spontaneous emission processes. These obviously affect the ideal fidelity of the gates implemented in the teleportation scheme. However  such unavoidable effects can be minimized by inserting  suitable quantum error correction codes in our scheme~\cite{Shor,Steane}. This corresponds to use more two-qubit gates for the same computation. Moreover, we expect that  in  an experimental setup the quantum teleportation process, would be repeated many times. In fact, the simulations performed in this work represent  a ``single shot'' of the network in Fig.~\ref{fig1},
with three electrons in the same minimum of the SAW. However, it is reasonable to think that in the experiment electrons also populate the other minima, as described in Ref.~\onlinecite{Barnes1}. This corresponds to a multiple repetition of the teleportation scheme (one each SAW minimum).

\section{Conclusions}  Here we have proposed  a device for
the deterministic teleportation of electrons injected and driven by
SAWs in a network of coupled quantum wires. It consists of a sequence
of single-qubit (beam splitter and phase shifter) and two-qubit
(Coulomb Coupler) gates which allows a high level of control over the
state evolution.  Numerical simulations show that, with a suitable
design of the nanostructure, the fidelity of the teleportation can
reach values close to 1, indicating  a high-realibility  of  the process.  Furthermore, we also mention the possibility of using carrier transport in edge states for quantum Hall effect regime, instead of SAW-assisted electron transport in quantum wires.

The experimental realization  of the device proposed is challenging  since it requires the use of frontier mesoscopic semiconductor technology. However,  both  the new developments in nanostructures fabrication  permitting the control of the coupling  between two modes of two 1D-channels~\cite{Fischer1,Fischer2}, and the recent observation of single electron dynamics in experiments  of  SAW assisted charge transport~\cite{Barnes2}, seem  to indicate the feasibility of an experimental setup of our device. Its realization would undoubtedly represent a great step forward toward quantum-computing capable architectures   scalable and integrable with traditional microelectronics.

\begin{acknowledgments}
We are pleased to thank Carlo Jacoboni for fruitful discussions.
We  acknowledge support from CNR
Progetto Supercalcolo 2008 CINECA.
\end{acknowledgments}

\end{document}